\documentclass{LT23auth}
\usepackage{graphicx}

\begin{document}

\begin{frontmatter}

\title{Magnetic Properties of a Two-Dimensional Mixed-Spin 
System}

\author[address1]{J.-H. Park\thanksref{thank1}},
\author[address2]{J. T. Culp},
\author[address3]{D. W. Hall},
\author[address2]{D. R. Talham},
\author[address1]{M. W. Meisel}

\address[address1]{Department of Physics and National High Magnetic Field 
Laboratory, University of Florida, Gainesville, FL  32611-8440, USA.}
\address[address2]{Department of Chemistry, University of Florida, Gainesville, 
FL  32611-7200, USA.}
\address[address3]{National High Magnetic Field Laboratory, Florida State 
University, Tallahassee, FL  32310, USA.}



\thanks[thank1]{Corresponding author. E-mail: juhyun@phys.ufl.edu}

\begin{abstract}
Using a Langmuir-Blodgett (LB) synthesis method, novel two-dimensional (2D) 
mixed-spin magnetic systems, in which each magnetic layer is both structurally
and magnetically isolated, have been generated.  Specifically, a 
2D Fe-Ni cyanide-bridged network with a face-centered square grid 
structure has been magnetically and structurally characterized.
The results indicate the presence of ferromagnetic exchange 
interactions between the Fe$^{3+}$ ($S=1/2$) and Ni$^{2+}$ ($S=1$) 
centers. 
\end{abstract}

\end{frontmatter}
\section{Introduction}
Part of the interest in molecule-based magnets arises from the potential for 
forming low-dimensional structures with anisotropic physical properties.
  In this paper, we report on the magnetic behavior of quasi-two-dimensional 
  (2D) Fe-CN-Ni grid networks, in which each magnetic layer is both
   structurally and magnetically isolated.  Culp \emph{et al.} have developed 
   a unique Langmuir-Blodgett (LB) synthesis method to transfer monolayers of
    a 2D Fe-CN-Ni grid network, formed at the air-water interface, as thin 
    films onto various substrates \cite{Culp}.
  This technique generates a 
    new low-dimensional network which cannot be formed from homogeneous
     reaction conditions.  
     
    The structure of the resulting assemblies of 
     mixed-spin Fe-CN-Ni bilayers transferred as thin films onto Mylar is 
     shown schematically in Fig. 1.  Each deposition cycle produces a bilayer 
     of the inorganic grid network sandwiched between an organic layer.
     The LB films have been structurally 
     characterized by x-ray photoelectron spectroscopy (XPS), FT-IR 
     spectroscopy, x-ray absorption fine structure (XAFS), and grazing 
     incidence synchrotron x-ray diffraction (GIXD).  These probes have
      allowed for the identification of a face-centered square grid structure
       with an Fe$^{3+}$$-$Ni$^{2+}$ separation of 5.1 \AA, and an average structural 
       domain size of 3600 \AA$^{2}$
\cite{Culp}.  Adjacent bilayers are separated by \linebreak 
$\approx$ 50 \AA, and are effectively 
isolated from each other.  
\begin{figure}[b]
\begin{center}\leavevmode
\includegraphics[width=0.8\linewidth]{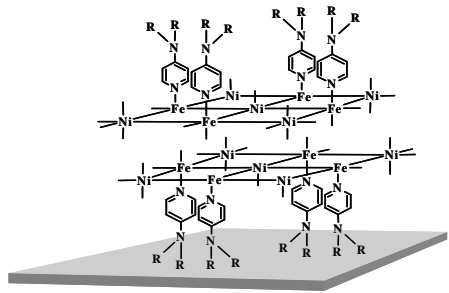}
\caption{The assembly of the Fe-CN-Ni grid bilayer transferred onto a thin mylar film.}
\label{Fig.1}\end{center}
\end{figure}

The existence of ferromagnetic exchange between the low spin Fe$^{3+}$ ($S=1/2$) 
and Ni$^{2+}$ ($S=1$) ions is expected after consideration of the exchange 
interaction in other cyanide bridged Fe$^{3+}$$-$Ni$^{2+}$ systems
 \cite{Gadet,Juszczyk}. 
 The ferromagnetic behavior in these 
 and other metal cyanides has been rationalized by considering the \linebreak
 d-orbital symmetries for the unpaired electrons in the respective metal centers. 
   The magnetic orbital on the Fe$^{3+}$ center is of t$_{2g}$ symmetry while 
   the two Ni$^{2+}$ spins are in orbitals 
    of e$_{g}$ symmetry.  The different symmetry orbitals do not overlap via the 
    cyanide ligand, thereby resulting in the parallel alignment of the spins according 
     to Hund's rule
\cite{Kahn}.  

\section{Magnetic Measurements}
The magnetic properties of a sample containing 150 bilayers of the 
Fe-CN-Ni network on each side of a Mylar film (10 cm$^{2}$) were investigated
 using SQUID magnetometry.  Measurements were performed with the applied
 magnetic field oriented either parallel or perpendicular to the plane of the
 film.  The field cooled (FC) magnetization
versus temperature for both orientations, measured in a field of 2 mT,
 are shown in Fig. 2.
       Both orientations show a break at a T$_{c}$ of 10 K, and the
       saturated behavior below 5 K indicates the presence of magnetically ordered
 domains.
\begin{figure}[t]
\begin{center}\leavevmode
\includegraphics[width=0.8\linewidth]{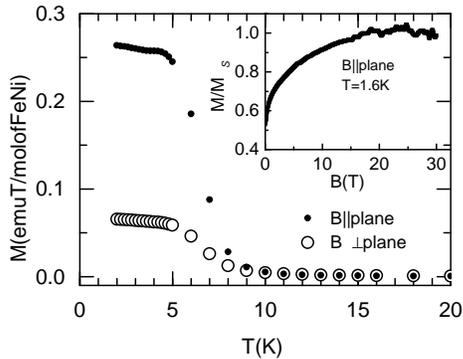}
\caption{Field cooled (FC) magnetization measured in 2 mT.  The high
field magnetization data at 1.6 K are shown in the inset.}
\label{Fig.2}\end{center}
\end{figure}
 The presence of long-range magnetic order at 2 K was 
verified by the measurements shown in Fig. 3.  The material shows clear hysteresis loops with 
coercive fields of 14 mT (parallel orientation) and 11 mT (perpendicular
 orientation).  
 
 The assignment of a ferromagnetic state was made based 
 on high field magnetization studies performed at the National High
  Magnetic Field Laboratory, Tallahassee (NHMFL).  The magnetization
   of the film was measured up to 
30 T using a vibrating sample magnetometer (VSM).  As shown in Fig. 2 (inset),
 the magnetization of the sample 
saturates around 20 T without evidence for a spin-flop transition.  The increased noise
 above 15 T is due to the addition
of a second dc power supply generating the magnetic field. The absence of a spin 
flop discriminates against 
antiferromagnetic coupling and the presence of a ferrimagnetic state. 
\section{Discussion and Summary}
A stronger magnetization measured in the parallel
  orientation in the M \emph{vs.} T plot in Fig. 2 indicates that the magnetic
   easy axis is within the XY directions parallel to the film 
   planes.  Consistently, in the low field limit (B $\ll$ B$_{S}$, where B$_{S}$
   is the applied filed at which M saturates, M$_{S}$), 
   the magnetization in the parallel orientation changes more rapidly 
   than in the perpendicular orientation with respect to changes 
   in the external field.  This behavior is most likely due to a 
   combination of the single ion anisotropies of the metal centers
    and the reduced symmetry of the low-dimensional material.
In summary, a 2D Fe-Ni cyanide-bridged network with a face-centered square grid 
structure has been generated using a Langmuir-Blodgett (LB) synthesis method, and the magnetic study
on this film shows the presence of ferromagnetism.
\begin{figure}[t]
\begin{center}\leavevmode
\includegraphics[width=0.8\linewidth]{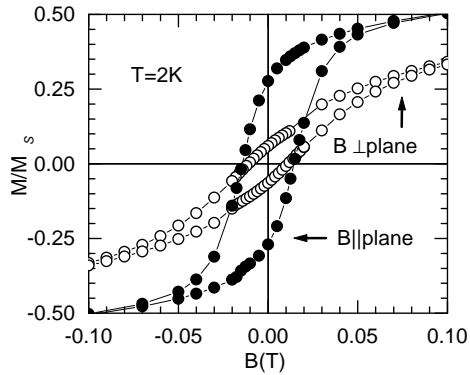}

\caption{M/M$_{S}$ \emph{vs.} B of the Fe-CN-Ni LB film in both orientations.  The measurement was
made at 2 K.}
\label{Fig.3}\end{center}
\end{figure}
\begin{ack}
This work was supported, in part, by NSF DMR-9900855, NSF DMR-0084173 
for the NHMFL, and ACS-PRF 36163-AC5.
\end{ack}
%



%

\end{document}